\documentclass[prl,aps,showpacs,twocolumn,superscriptaddress,floatfix]{revtex4}

\usepackage{graphicx}
\usepackage{epsfig}
\usepackage{amsfonts}
\usepackage{amsmath,amssymb}
\usepackage{bm}

\newcommand{\gras}[1]{\boldsymbol{#1}}

\begin{document}

\title{Charge radii and neutron correlations in helium halo nuclei}%

\author{G. Papadimitriou}
\affiliation{
Department of Physics and Astronomy, University of Tennessee, Knoxville, Tennessee 37996, USA
}%
\affiliation{
Physics Division, Oak Ridge National Laboratory, Oak Ridge, Tennessee 37831, USA
}%
\author{A.T. Kruppa}
\affiliation{
Department of Physics and Astronomy, University of Tennessee, Knoxville, Tennessee 37996, USA
}%
\affiliation{Institute of Nuclear Research,
  P.O. Box 51, H-4001 Debrecen, Hungary}

\author{N. Michel}%
\affiliation{Department of Physics, Post Office Box 35 (YFL), FI-40014 University of Jyv\"{a}skyl\"{a}, Finland}

\author{W. Nazarewicz}
\affiliation{
Department of Physics and Astronomy, University of Tennessee, Knoxville, Tennessee 37996, USA
}%
\affiliation{
Physics Division, Oak Ridge National Laboratory, Oak Ridge, Tennessee 37831, USA
}%
\affiliation{
Institute of Theoretical Physics, University of Warsaw, ul. Ho\.za 69,
PL-00-681 Warsaw, Poland }%

\author{M. P{\l}oszajczak}
\affiliation{Grand Acc\'{e}l\'{e}rateur National d'Ions Lourds (GANIL), CEA/DSM-CNRS/IN2P3, BP 55027, F-14076 Caen Cedex, France}

\author{J.Rotureau}
\affiliation{Department of Physics, University of Arizona, Tucson, Arizona 85721, USA}


\begin{abstract}
Within  the complex-energy configuration interaction framework, we study   correlations of valence neutrons to explain the behavior of charge radii in the neutron halo nuclei $^{6,8}$He.
We find that the experimentally observed decrease of the charge radius between $^6$He and $^8$He  is caused by a subtle interplay between three effects: dineutron correlations, a spin-orbit contribution to the charge radius, and a core swelling effect. We demonstrate that two-neutron angular correlations in the $2^+_1$ resonance of $^6$He  differ markedly from the ground-state correlations
in $^{6,8}$He. Finally, we discuss the impact of  the neutron threshold position and valence neutron correlation energy on the neutron radius, i.e., the pairing-antihalo effect.
\end{abstract}


\maketitle



\textit{Introduction}---The physics of Open Quantum Systems (OQS) \cite{OQSbook1} has attracted a lot of attention in many fields of physics, including  atomic and molecular physics, quantum optics, condensed matter physics,  nuclear physics, and tests of quantum mechanics. In atomic nuclei, the ``openness" of the system manifests itself by the coupling to the  many-body continuum representing various decay and reaction channels \cite{Okolowicz,N.M4}. In addition to being  prototypical  OQSs, nuclei   are excellent laboratories of many-body physics. While the number of fermions  in  nuclei is very small compared to atoms, molecules, and solids, nuclei exhibit an emergent behavior that is present in other complex systems. Due to the presence of particle thresholds,
 atomic nuclei form a network of
correlated fermionic systems interconnected via reaction channels. The effects  due to openness are enhanced  in weakly bound nuclei \cite{witek_jacek}. Indeed,
an essential part of the motion of those short-lived systems, such as extended nuclear halos, is in classically forbidden regions, and their properties are profoundly impacted by both the continuum and many-body correlations.

The neutron-rich helium isotopes $^{6}$He and $^{8}$He, which are subjects of this study, are excellent representatives of OQS. Indeed, both nuclei  are Borromean halos, they have no bound excited states, and they exhibit the binding-energy anomaly, i.e., the presence of higher one- and two- neutron emission thresholds in  $^{8}$He  than in  $^{6}$He.
Studies of charge radii of nuclear  halos represent a splendid example of the cross-fertilization between
atomic and nuclear physics in the field  of OQSs. Experimentally, charge radii of helium halos
were extracted from measured isotopic shifts of helium atoms in
pioneering studies of Refs.~\cite{L.B.Wang,Mueller}. Very recently, these numbers  were revaluated \cite{Brodeur} based on precision mass measurements. It was found that the charge radius of $^6$He, 2.059(7) fm, exceeds that of $^8$He, 1.959(16) fm, and both radii  are much larger than the charge radius of $^4$He (1.681(4) fm \cite{I.Sick}). Since the charged protons are confined inside the tightly bound $\alpha$-core, the differences of
charge radii of helium halos carry unique structural information on nuclear Hamiltonians and nuclear many-body dynamics.

Because of their precision, charge radius data provide a critical test  of nuclear models. Charge radii have been calculated successfully within the {\it ab initio}
GFMC \cite{Pieper_Wiringa} and NCSM \cite{Navratil} frameworks
using state-of-the-art realistic interactions.
However, not much is known about the particle correlations that cause this behavior. To obtain a simple physical picture of the observed effect and
understand the  physics that governs the neutron distributions in helium halos,
in this Letter, we employ the complex-energy continuum shell model
\cite{N.M4}, the
Gamow Shell Model (GSM) \cite{N.M1}. GSM is a configuration interaction
approach with a single-particle (s.p.) basis given
by the Berggren ensemble \cite{Berggren}, which consists of Gamow
(bound and resonance) states and the nonresonant scattering
continuum. One can find successful applications of the Berggren ensemble to the helium halos in the earlier GSM calculations \cite{N.M1,Michel_mirror,Hagen_Michel} and also in \textit{ab initio} coupled-cluster calculations of Ref. \cite{Gaute1}.
Since the dynamics of halo neutrons in $^{6,8}$He is primarily governed by many-body correlations and continuum coupling, GSM is the tool of choice
to get a simple insight into underlying physics.

\textit{Model}---We assume  that the halo nucleus can be
described as a system of $n_\nu$ valence
neutrons moving around a closed $^4$He core.
To cope with the problem of spurious center-of-mass motion, we adopt a system of intrinsic nucleon-core
coordinates inspired by the cluster orbital shell model \cite{Ikeda}.
In these coordinates, the translationally invariant
GSM Hamiltonian is:
%
$H= \sum_{i=1}^{n_\nu}\left [ \frac{p_{i}^{2}}{2\mu} + U_{i} \right] + \sum_{i<j}^{n_\nu} \left[ V_{ij} + \frac{1}{A_{c}} \gras{p}_{i}\gras{p}_{j} \right]$,
%
where $U_i$ is the single-particle (s.p.) potential describing the field of the
core, $V_{ij}$ is the two-body residual interaction between
valence neutrons, and the last term represents the two-body energy recoil with $A_c=4$ being the mass of the core.
For the one-body potential, we took the ``$^5$He" Woods-Saxon (WS) field (with spin-orbit term)  of Ref.~\cite{N.M1} which reproduces
the experimental energies and widths of known s.p. 3/2$^-$ and 1/2$^-$ resonances in $^5$He.
The residual interaction employed is a finite-range Minnesota (MN) central  potential \cite{Tang}, which is a sum of three
Gaussians with different ranges. Since only a $T$=1 interaction channel is present,  two Gaussians are sufficient; this has been achieved by  taking  the parameter $u$=1 of the original MN potential \cite{Tang}. As the interaction coupling constants should depend on the configuration space used,
in our work we alter the Gaussian strengths $V_{0R}$ and $V_{0s}$ of the MN potential
to reproduce the experimental ground state (g.s.) energies of $^{6,8}$He
($V_{0R}=250.2$\,MeV and $V_{0s}=-110.1$\,MeV); the Gaussian
ranges are the same as in Ref.~\cite{Tang}.
The one-body basis is given by the Berggren ensemble, which for the case of $^6$He  is that of  the ``$^5$He" WS field.
For the heavier helium isotopes,  the quality of
the WS basis deteriorates; hence, we use a  Gamow Hartree-Fock (GHF) ensemble  \cite{N.M3}  generated by the
WS potential and the modified  MN interaction.
The two-body  matrix elements of $V_{ij}$
and the recoil term  are efficiently computed by using the Harmonic Oscillator (HO) expansion method \cite{Hagen_Michel,Michel_mirror}.
This requires  calculating the overlaps between Berggren basis and HO states.
Because of the Gaussian asymptotic behavior of HO wave functions, no complex scaling is needed, since these overlaps always converge.
By taking the HO length parameter  $b$=2\,fm and the lowest 16 HO states for each partial wave, we obtain full convergence and perfect agreement with
the Complex Scaling (CS) approach in the Slater basis \cite{Kruppa_et_al}.
The  calculations were carried out in a large shell model space consisting of the 0$p_{3/2}$ Gamow resonance and non-resonant $psd$ scattering continua.
The maximum s.p. momentum was chosen to be $k_{max} = 4.0$\,fm$^{-1}$.
The $p_{3/2}$ contour in the complex momentum plane  was discretized with a total of 30 points, while the remaining non-resonant continua were chosen along  the real momentum axis and discretized with  21 points each; this resulted
 in a large GSM space of 115 shells.
In this space,  the dimension of the GSM Hamiltonian matrix for
the $^8$He becomes prohibitively large, and, for this reason, we applied the Density Matrix Renormalization Group approach earlier adopted to GSM in Refs.~\cite{Jimmy}. Since the $s$-wave enters the Berggren ensemble, in order to satisfy the Pauli Principle between core and valence particles we apply the Saito's Orthogonality Condition Model \cite{Saito}.

\textit{Radii and neutron correlations}---To obtain the charge radii of $^{6,8}$He we first express the proton point radius  in the intrinsic set of
coordinates: $\langle r^{2}_{pp}(^{A_{c}+n_\nu}X) \rangle = \langle r^{2}_{pp}(^{A_{c}}X) \rangle +
  \frac{1}{(A_{c}+n_\nu)^{2}}\sum_{i=1}^{n_\nu}\langle r_{i}^{2} \rangle +
  \frac{2}{(A_{c}+n_\nu)^{2}}\sum_{i<j}^{n_\nu}\langle \gras{r}_{i}\cdot  \gras{r}_{j}\rangle$, where
the first term is the point proton radius of the $\alpha$-core
in nucleus $^{A_{c}+n_\nu}X$, and the two remaining terms represent
the contribution to the proton radius from  the  motion of the core around the
nuclear center-of-mass, i.e., recoil effect.
The {\it ab initio} GFMC calculations  \cite{Pieper_Wiringa2} predict the 4.58\% and 6.66\% increase  of the $\alpha$-particle
proton point radius in $^6$He and $^8$He, respectively. (The effect of core polarization by valence neutrons cannot be  totally decoupled from the effect of  $\gras{\tau}\cdot\gras{\tau}$ forces, which virtually exchange protons with neutrons.) By assuming these numbers, together with the experimental value of 1.46\,fm for the $\alpha$-particle point radius, we adopt
the point proton radius of  1.527\,fm (1.557\,fm) for $^6$He ($^8$He).
This core ``swelling" effect due to valence neutrons
is large enough that it cannot be  neglected in the detailed analysis.
To make contact with the measured charge radii, we correct
the calculated point-proton radii for the finite sizes of proton and neutron through the usual expression \cite{Friar_Negele}: $\langle r_{ch}^{2} \rangle =  \langle r_{pp}^{2} \rangle + \langle R_{p}^{2} \rangle + \frac{N}{Z}\langle R_{n}^{2} \rangle + \frac{3}{4M_{p}^{2}} + \langle r^{2} \rangle_{so}$,
where $\langle R_{p}^{2} \rangle=0.769$\,fm$^2$  (proton charge radius),
$ \langle R_{n}^{2} \rangle=-0.1161$\,fm$^2$ \cite{Yao} (neutron charge radius),  $\frac{3}{4M_{p}^{2}}=0.033$\,fm$^2$ \cite{Sprung_Marto} (Darwin-Foldy term), and $\langle r^{2} \rangle_{so}$ is the spin-orbit (s.o.) contribution.

The correlations amongst the valence neutrons are assessed through the two-body  density
$\rho_{nn}(r,r^{\prime},\theta) = \langle \Psi |\delta(r_1-r)\delta(r_2-r^{\prime})\delta(\theta_{12}-\theta)|\Psi\rangle$, with $r_1$ ($r_2$) being the distance between the core and the
first (second) neutron and $\theta_{12}$ - the opening angle between the
two neutrons. The density $\rho_{nn}(r,r^{\prime},\theta)$
differs from the two-particle density
of Refs.~\cite{Bertsch_pair,Sagawa_2005} by the absence of the
Jacobian $8\pi^2 r^2 r'^2 \sin\theta$. Consequently,
$\int\rho_{nn}(r,r^{\prime},\theta) drdr'd\theta = 1.$
In practical implementation, $\rho_{nn}(r,r',\theta)$ has been expressed in the $LS$ coupling scheme \cite{Sagawa_2005} and benchmarked against  CS calculations \cite{Kruppa_et_al}.

\textit{Results}---The adopted GSM Hamiltonian reproduces the energetics of the  isotopic chain $^{5-8}$He, in particular the one-neutron (1n) and two-neutron (2n) thresholds. We predict  the 2$^+_1$ resonance in $^6$He
to have  energy 851\,keV ($E_{exp}=822$\,keV \cite{nuclear_dat})
and width $\Gamma = 109$\, keV ($\Gamma_{exp}=113$\,keV). The predicted  3/2$^-$  g.s. resonance of $^7$He has $E= -0.707$\, MeV ($E_{exp}=-0.528$\,MeV).

\begin{figure}[h!] 
  \includegraphics[width=\columnwidth]{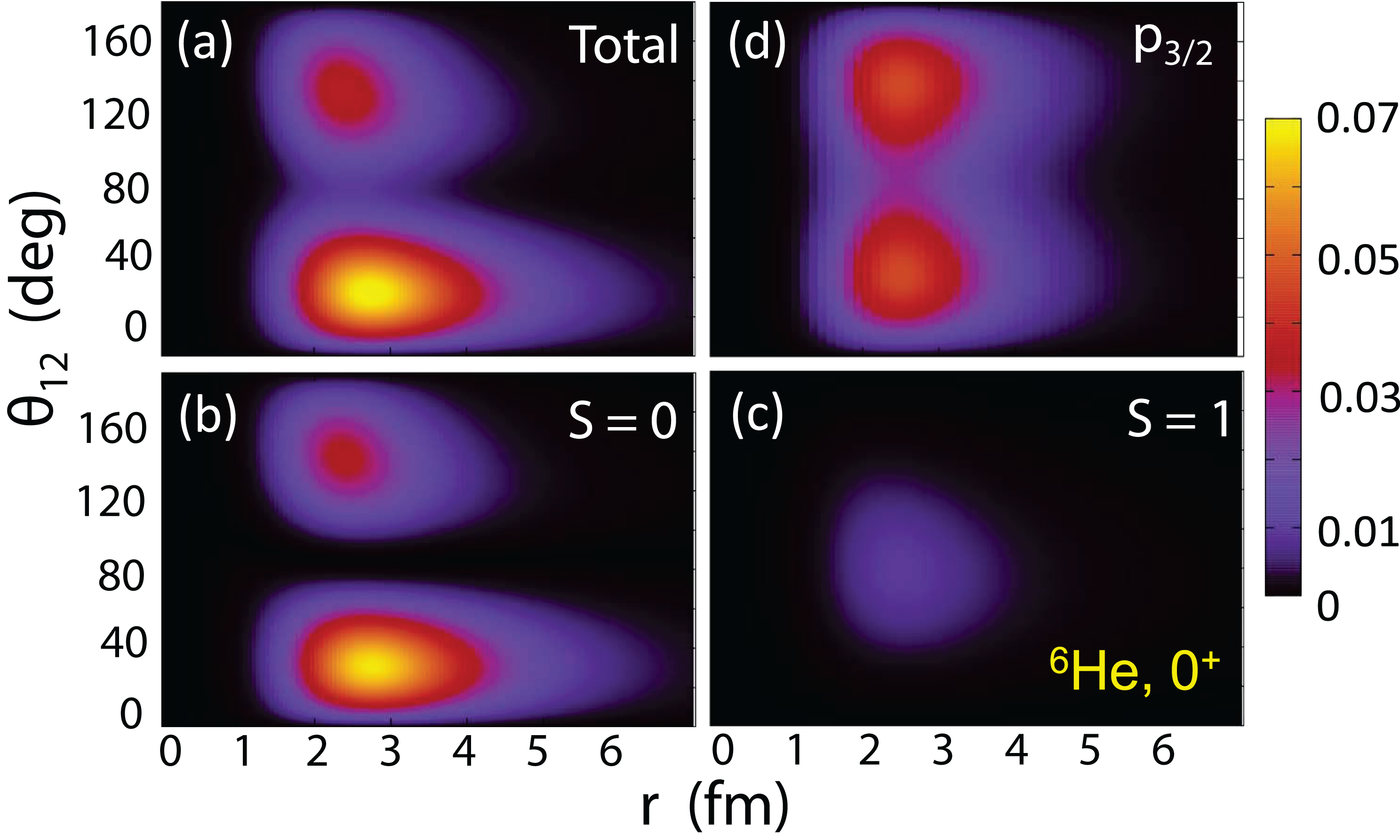}
  \caption[T]{\label{fig1}
  (Color online) Two-neutron GSM density (in fm$^{-2}$) of the 0$^+$ ground state of $^6$He: (a) total; (b) $S=0$ component; (c) $S=1$ component; (d) density in a $p_{3/2}$  model space.}
\end{figure}

Before discussing GSM predictions for the charge and neutron radii, we present
results for 2n correlations. Figure~\ref{fig1}(a) shows the GSM density
$\rho_{nn}(r,\theta)=\rho_{nn}(r_1=r,r_2=r,\theta)$ for the g.s. of $^6$He.
The density exhibits two peaks \cite{Vaagen,Oganessian,Sagawa_2005,Kikuchi}. One maximum, corresponding to a small opening angle and a large radial extension, represents the dineutron configuration. The second maximum, found in the region of large angles and radially well localized, represents the cigar-like configuration. Similar to Ref.~\cite{Sagawa_2005}, we find that both configurations have a dominant  $S=0$ component in which the two neutrons are in the spin singlet state (see Fig.~\ref{fig1}(b)). The amplitude of the $S=1$ density component shown in Fig.~\ref{fig1}(c) is fairly small (13\%); this contribution to density has a very broad maximum around $\theta_{12}=90^\circ$.
To illustrate  the importance of the continuum coupling for the dominance of the dineutron configuration in   $^6$He, in Fig.~\ref{fig1}(d)  we plot
$\rho_{nn}(r,\theta)$ calculated in the limited $p_{3/2}$ model space.
Interestingly, the omission of $p_{1/2}sd$ scattering continua, which enter  the g.s. wave function of $^6$He with fairly small amplitudes \cite{Michel_mirror},  has  a rather dramatic effect on 2n  distribution: the dineutron component loses its halo character, and the heights of the dineutron and cigar-like peaks become equal.

The unique feature of GSM is that it  enables us to examine many-body correlations in unbound states. To this end, we study $\rho_{nn}$
in the  $2^+_1$ state of $^6$He, a 2n resonance.
\begin{figure}[t]
  \includegraphics[width=\columnwidth]{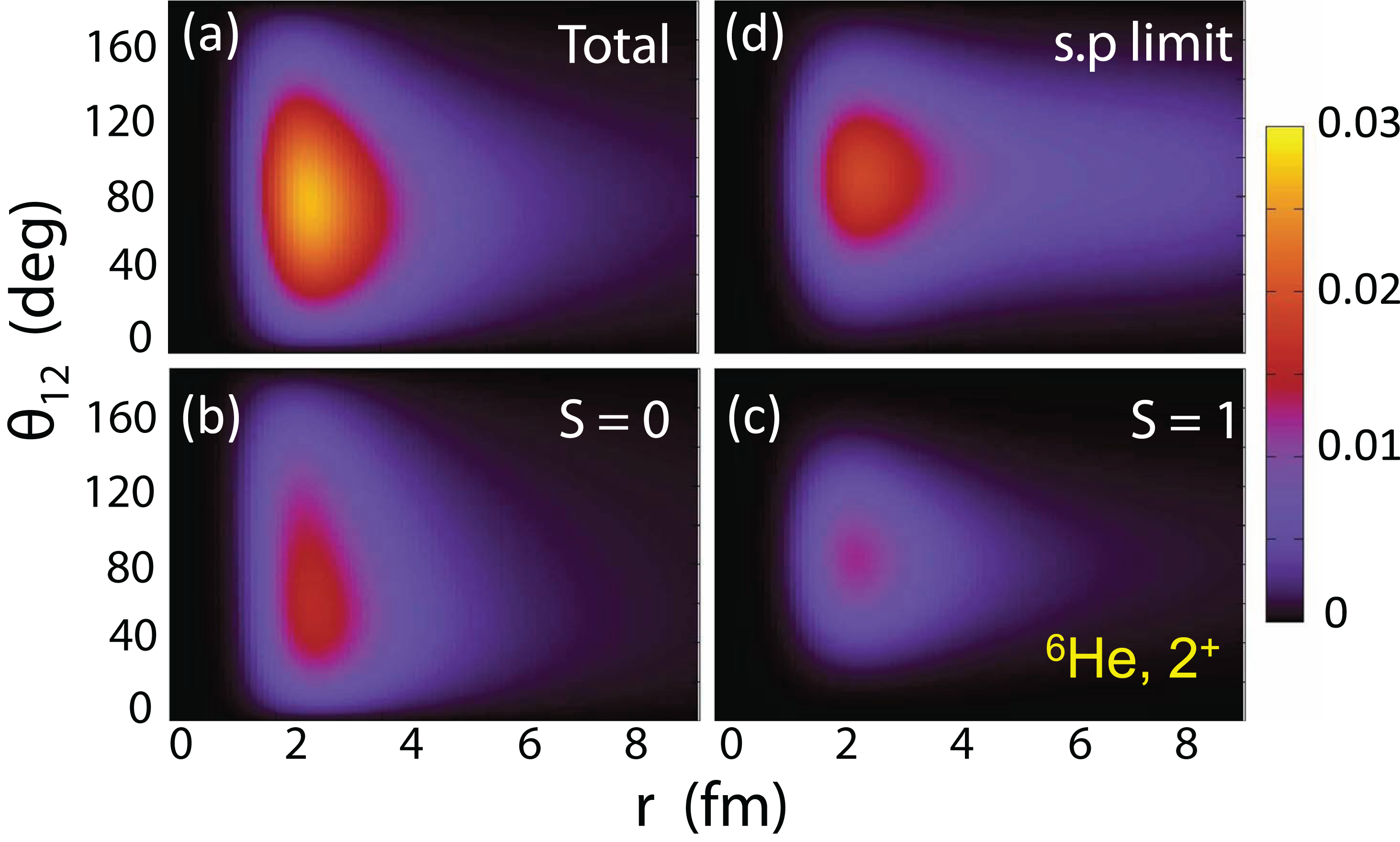}
  \caption[T]{(Color online)  Two-neutron GSM density (in fm$^{-2}$) of the 2$^+_1$ resonance in $^6$He:  (a) total; (b) $S=0$ component; (c) $S=1$ component; (d) density with the two-body interaction turned off. }\label{fig2}
\end{figure}
As seen in Fig.~\ref{fig2}(a), the corresponding 2n density is radially extended, which nicely illustrates the unbound character of the 2$^+$ state. Contrary to the g.s. density, $\rho_{nn}(r,\theta)$
is characterized by one broad maximum centered around $\theta_{12}=60^\circ$.
We find  that  the
$S=0$ and $S=1$ configurations have similar spatial distributions, and
 the ratio of their amplitudes is $\sim$2:1. The GSM wave function of
the 2$^+$ resonance is dominated by a $(0p_{3/2})^2$ resonant component \cite{Michel_mirror}. To demonstrate the s.p. character of this state, in
Fig.~\ref{fig2}(d) we plot $\rho_{nn}(r,\theta)$ obtained by turning  the residual interaction off, thus  preventing scattering from the $0p_{3/2}$ Gamow state to the non-resonant continua. We see that the resulting density is close to that in the full space. This result suggests that the valence neutrons in the first excited state of $^6$He are weakly correlated.

\begin{figure}[h!] 
  \includegraphics[width=0.77\columnwidth]{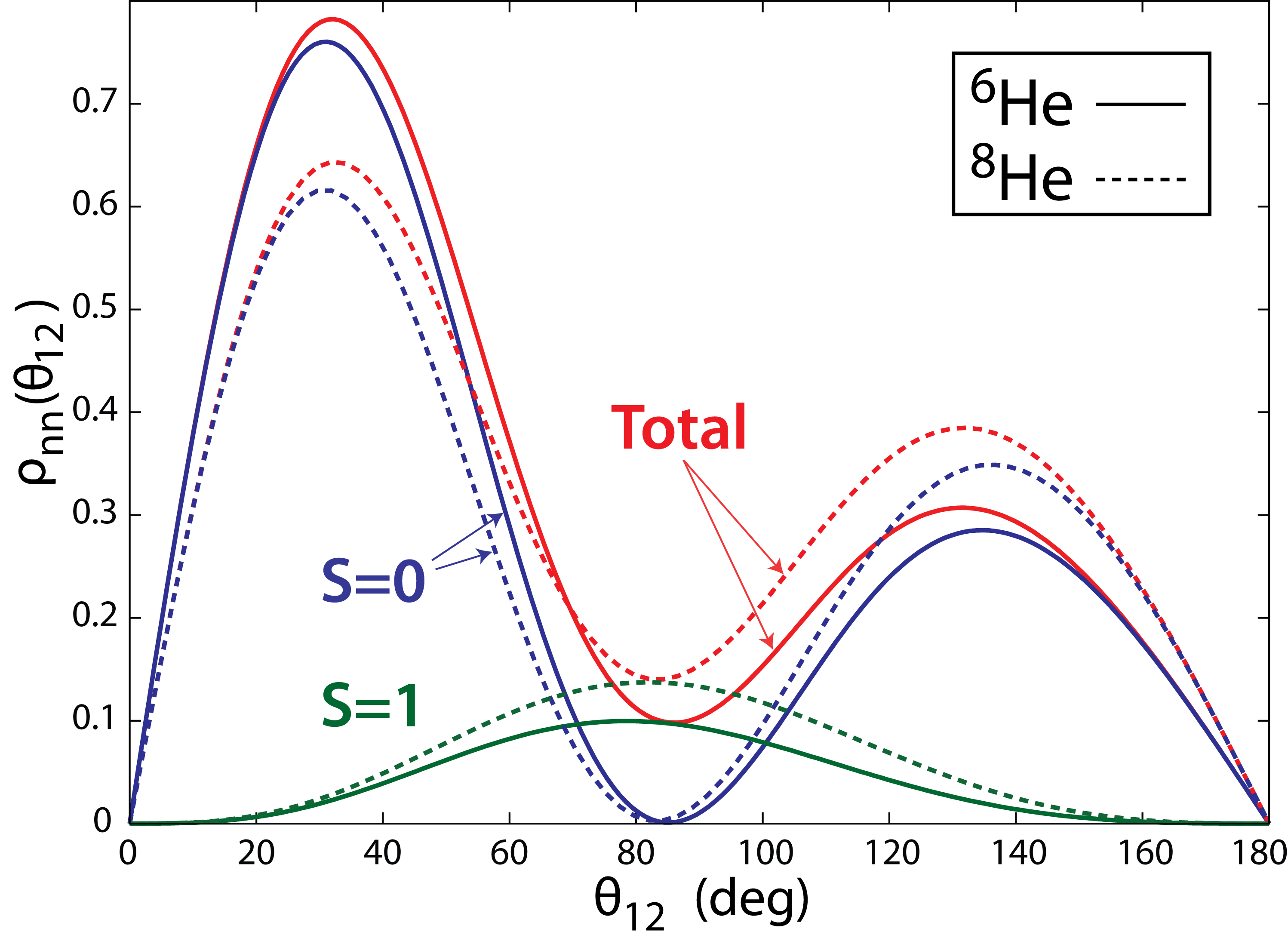}
  \caption[T]{(Color online) Angular 2n GSM densities ($S$=0, 1, and total) for g.s. configurations of  $^{6}$He (solid lines) and $^{8}$He (dashed lines).}\label{fig3}
\end{figure}

We next study the neutron correlations in g.s. of $^8$He. Since $\rho_{nn}(r,\theta)$ in $^6$He and $^8$He are fairly similar, to better see the difference between these two cases,  in Fig.~\ref{fig3} we compare the angular correlation densities
$\rho_{nn}(\theta_{12}) = \int \,dr_1 \int \,dr_2 \rho_{nn}(r_1,r_2,\theta_{12})$.
Our calculations demonstrate that the weight of the dineutron structure in $^8$He is less than in $^6$He due to increased strengths of cigar-like and $S=1$ components. The addition of the two extra neutrons
results in a slight increase in the average opening angle from $\theta_{12} = 68^\circ$ in $^6$He  to $\theta_{12} = 78^\circ$ in $^8$He.
It is worth noting, however, that the  positions of the two peaks in $\rho_{nn}(\theta_{12})$  are practically the same in $^{6,8}$He, and the 2n correlation density does not  broaden up (becomes more ``democratic") in $^8$He.

\begin{figure}[h!] 
  \includegraphics[width=0.9\columnwidth]{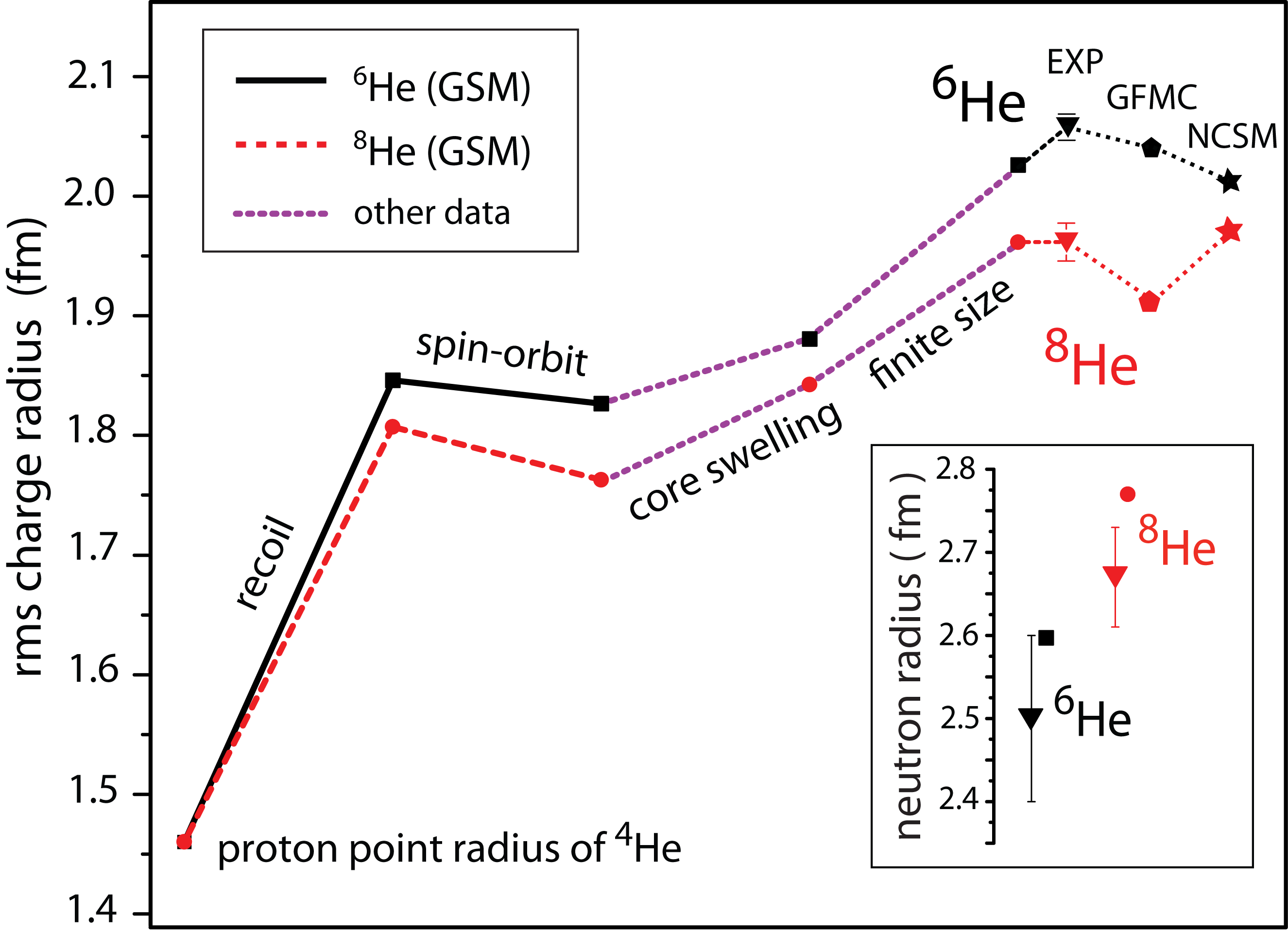}
  \caption[T]{(Color online) Different contributions to the charge radius of $^{6}$He (solid line, squares) and $^{8}$He (dashed line, dots) calculated in GSM. The core swelling contribution is taken from GFMC calculations of  Ref.~\cite{Pieper_Wiringa2}. Recently revised experimental charge radii come from \cite{Brodeur} (triangles).
 The  NCSM \cite{Navratil} (stars) and GFMC \cite{Pieper_Wiringa} (pentagons) results are marked for comparison. The inset shows GSM rms neutron radii  compared to experiment
 \cite{Alkhazov}.}\label{fig4}
\end{figure}

It has recently been pointed out \cite{Ong}
that $\langle r^2 \rangle_{so}$ may give
an appreciable contribution to the charge radii of halo nuclei. We compute the s.o. correction  in GSM following Ref.~\cite{Friar_Negele}.
The calculated s.o. rms radius of $^6$He ($^8$He)
is found to be $\langle r^2 \rangle_{so} = -0.0718 fm^{2}$
($-0.158 fm^{2}$). These values are not very different from s.p. estimates of Ref.~\cite{Ong}; namely, $-0.08 fm^2$  and $-0.17 fm^2$. The final results for charge radii are displayed in Fig.~\ref{fig4}: the smaller core recoil  and larger s.o. effect in $^8$He both contribute to the reduced value of $\sqrt{\langle r^{2}_{ch} \rangle}$ in this nucleus as compared to $^6$He. It is to be noted that
the recoil effect is primarily sensitive to the 2n threshold energy
and the position of the Gamow $0p_{3/2}$ pole \cite{GeorgePRC}.
The predicted total  values,
2.026 fm for $^6$He and 1.961 fm for $^8$He, are in nice agreement with experiment and {\it ab initio} GFMC and NCSM results.
Also, our predicted values of rms neutron radii, shown in the inset of Fig.~\ref{fig4}, are consistent with experimental data  \cite{Alkhazov}.

\begin{figure}[h!] 
  \includegraphics[width=0.9\columnwidth]{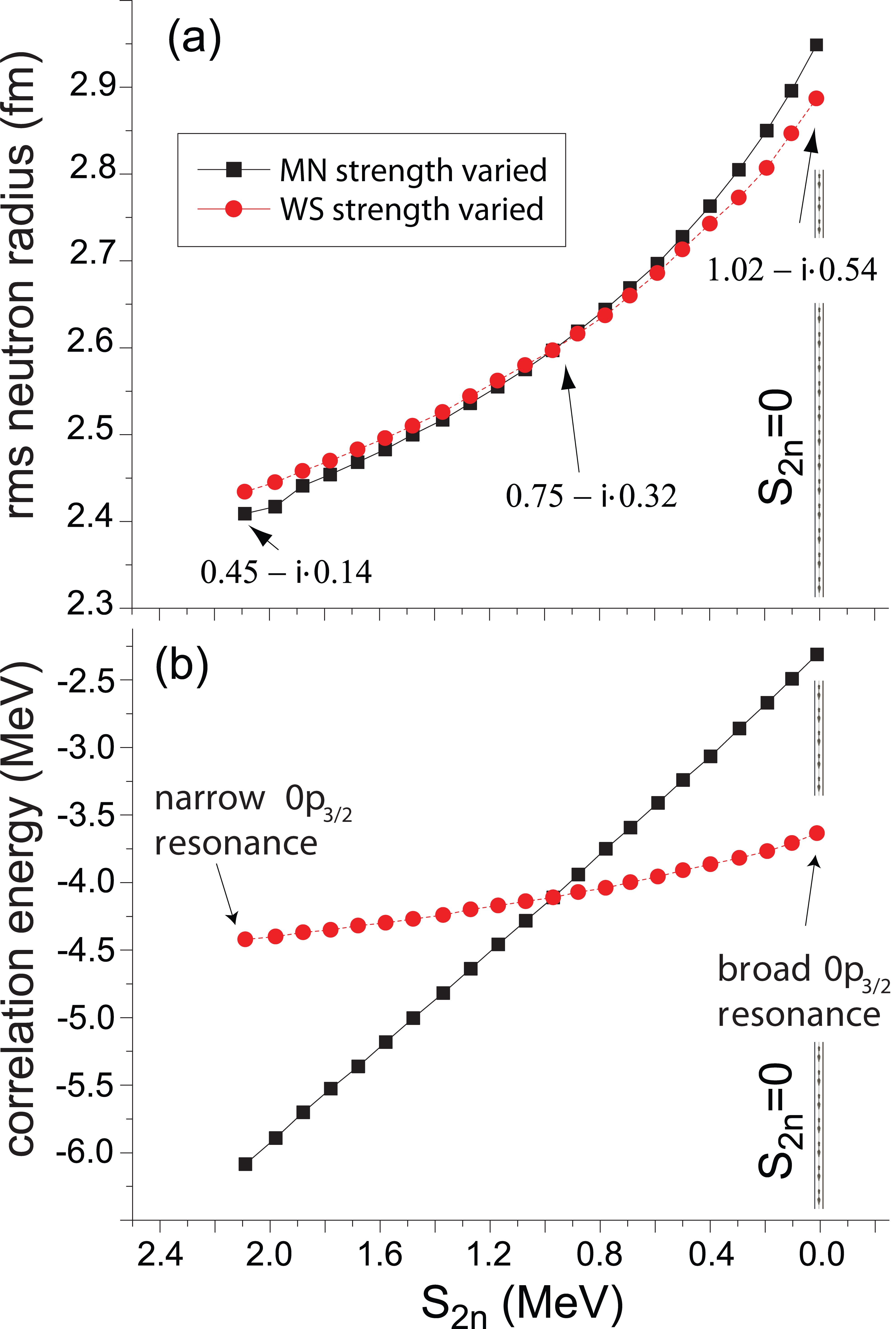}
   \caption[T]{(Color online) Top: Evolution of the GSM rms neutron radius of $^6$He as a function of the 2n separation energy $S_{2n}$. The position of the  2n threshold is varied by  (i) changing the strength of the MN interaction  (squares) or by (ii)  changing the strength of the WS potential (circles).
The complex energy $E-i\Gamma/2$ (in MeV) of the $0p_{3/2}$ Gamow resonant state
of the WS potential is marked for three values of $S_{2n}$.
Bottom: Correlation energy, i.e., the expectation value of the MN interaction,
in the cases (i) and (ii). The 2n threshold $S_{2n}=0$ is marked by a vertical line.
}\label{fig5}
\end{figure}
It is well known  \cite{witek_jacek} that in the vicinity of the particle threshold  pairing correlations
can profoundly modify properties of the system. One example is the Pairing-Anti Halo (PAH) effect \cite{Karim,Yamagami,Yamagami2,Rotival,Hagino_PAH}, in which pairing correlations in the weakly-bound even-particle  system change the asymptotic behavior of particle density
thus reducing its radial extension. To assess the sensitivity of the valence neutron extension on the position
of the 2n threshold,
in Fig.~\ref{fig5} we vary the 2n separation
energy S$_{2n}$ of $^6$He by either changing the $V_{0s}$ singlet
 strength of the  MN interaction that controls the amount of pairing correlations between the valence neutrons (variant 1; V1)  or by changing the  depth of the WS potential $V_{\rm WS}$ that determines the position of the crucial  $0p_{3/2}$ Gamow pole (variant 2; V2).
The neutron correlation energy can be estimated by calculating the expectation value of the MN interaction. It is seen in Fig.~\ref{fig5}(b) that
this quantity strongly depends on $V_{0s}$ in V1 and weakly on $V_{\rm WS}$
in V2. The rms neutron radius of $^6$He displayed in Fig.~\ref{fig5}(a)
gradually increases when approaching the 2n threshold $S_{2n}=0$ in both variants. The faster increase of neutron radius in V1 reflects the more rapid decrease of neutron pairing correlations with $S_{2n}$ in this case.
The character of the $0p_{3/2}$ resonance in V2 gradually changes from a moderately narrow one at $S_{2n} =2$\,MeV to very broad ($\Gamma \approx 1$\,MeV) at $S_{2n} =0$. In spite of this, the neutron radius gradually approaches the 2n threshold without exhibiting the rapid increase characteristic of an odd-particle halo system \cite{Karim}. Indeed, in the absence of correlations, the  rms radius  of the $0p_{3/2}$ state is expected to diverge as $S_{1n}^{-1/2}$ when approaching the threshold \cite{[Rii92]}. In summary, the result presented in Fig.~\ref{fig5} is a manifestation of the PAH effect in the GSM.

\textit{Conclusions}---We have applied the translationally invariant GSM with finite-range modified MN interaction to valence neutron correlations, and neutron and charge radii of the halo nuclei $^{6,8}$He. Our s.p. basis consists of the 0$p_{3/2}$ Gamow resonance and non-resonant $psd$ scattering continua  resulting in a large configuration space.
We obtain good agreement with experiment for charge and neutron rms radii, neutron separation energies in the He chain, and the $2^+_1$ resonance in $^6$He.
We find that
the charge radii of helium halos depend mainly on three factors: (i)  valence neutron correlations  resulting in a core recoil; (ii) s.o. contribution to the charge radius,  and (iii)  polarization of the core by valence neutrons.
We demonstrated
that the reduction of the charge radius when going from $^6$He  to $^8$He is
not due to a  more ``democratic" arrangement of the neutrons around the core but
rather due to  the reduction of the amplitude of the dineutron configuration
in the g.s. wave function, resulting in a smaller core recoil radius. In addition, the negative  s.o. radius contribution doubles with the addition of two valence neutrons. The dineutron  configuration
in $^{6,8}$He is strongly enhanced by coupling to the nonresonant continuum.
We studied  2n correlations in the $2^+_1$  resonance of $^6$He and found a rather broad distribution characteristic of an uncorrelated s.p. motion with no dineutron component. Finally, we demonstrated the presence of the PAH effect in the rms neutron radius of $^6$He when approaching the 2n threshold. This is the first evidence for this phenomenon in a configuration-interaction-based framework.

Useful discussions with J. Friar, S. Pieper, and  B. Wiringa
are gratefully acknowledged. This work was supported  by the Office of
Nuclear Physics,  U.S. Department of Energy under Contract No.
DE-FG02-96ER40963, by  the Hungarian OTKA Fund No. K72357, and  by the Academy of Finland and University of Jyv\"äskyl\"ä within the FIDIPRO programme.

%
%

\end{document}